\documentclass[aps,prb,reprint,showpacs,amsmath,amssymb,superscriptaddress]{revtex4-1}

\usepackage{graphicx}
\usepackage{ulem}
\usepackage{dcolumn}
\usepackage{bm}

\begin{document}

\title{Model of bound interface dynamics for coupled magnetic domain walls}

\author{P.~Politi}
\email{paolo.politi@isc.cnr.it}
\affiliation{Istituto dei Sistemi Complessi, Consiglio Nazionale delle Ricerche, Via Madonna del Piano 10, 50019 Sesto Fiorentino, Italy.}

\author{P. J.~Metaxas}
\email{metaxas@physics.uwa.edu.au}
\affiliation{School of Physics, M013, University of Western Australia, 35 Stirling Hwy, Crawley WA 6009, Australia.}
\affiliation{Laboratoire de Physique des Solides, Universit\'e Paris-Sud, CNRS, UMR 8502, 91405 Orsay Cedex, France.}
\affiliation{Unit\'e Mixte de Physique CNRS/Thales, 1 Avenue Augustin Fresnel, 91767 Palaiseau Cedex, France and Universit\'e Paris-Sud 11, 91405, Orsay Cedex, France.}

\author{J.-P. Jamet}
\affiliation{Laboratoire de Physique des Solides, Universit\'e Paris-Sud, CNRS, UMR 8502, 91405 Orsay Cedex, France.}

\author{R. L.~Stamps}
\affiliation{School of Physics, M013, University of Western Australia, 35 Stirling Hwy, Crawley WA 6009, Australia.}
\affiliation{SUPA--School of Physics and Astronomy, University of Glasgow, G12 8QQ Glasgow, United Kingdom.}

\author{J.~Ferr\'{e}}
\affiliation{Laboratoire de Physique des Solides, Universit\'e Paris-Sud, CNRS, UMR 8502, 91405 Orsay Cedex, France.}

\date{\today}

\newcommand{\be}{\begin{equation}}
\newcommand{\ee}{\end{equation}}
\newcommand{\bea}{\begin{eqnarray}}
\newcommand{\eea}{\end{eqnarray}}
\newcommand{\bse}{\begin{subequations}}
\newcommand{\ese}{\end{subequations}}
\newcommand{\bau}{\hbox{$\bar a_1$}}
\newcommand{\bad}{\hbox{$\bar a_2$}}
\newcommand{\bai}{\hbox{$\bar a_i$}}
\newcommand{\bab}{\hbox{$\bar a_b$}}
\newcommand{\Hdep}{H_{\hbox{\tiny dep}}}
\newcommand{\Ms}{M_{\hbox{\tiny S}}}
\newcommand{\vc}{v_{\hbox{\tiny C}}}
\renewcommand{\sf}{^{\hbox{\tiny{sf}}}}

\begin{abstract}
A domain wall in a ferromagnetic system will move under the action of an external magnetic field.
Ultrathin Co layers sandwiched between Pt have been shown to be a suitable experimental
realization of a weakly
disordered 2D medium in which to study the dynamics of 1D interfaces (magnetic domain walls). The behavior of these systems is encapsulated
in the velocity-field response $v(H)$ of the domain walls.
In a recent paper [P.J. Metaxas {\it et al.}, Phys. Rev. Lett. {\bf 104}, 237206 (2010)]
we  studied the effect of ferromagnetic coupling between two such ultrathin layers, each exhibiting different $v(H)$
characteristics. The main result was the existence of bound states over finite-width field ranges, wherein walls in the two layers moved
together at the same speed. Here, we discuss in detail the theory of domain wall dynamics in coupled
systems. In particular, we show that a bound creep state is expected for vanishing $H$ and
we give the analytical, parameter free expression for its velocity which agrees well with
experimental results.
\end{abstract}

\maketitle


\section{Introduction}

A number of physical phenomena involve elastic interfaces moving through disordered media. These phenomena range from domain wall motion in ferromagnets \cite{Lemerle1998,Repain2004epl,Metaxas2007,Kim2009}, ferroelectrics \cite{Paruch2006} and multiferroics \cite{Catalan2008} to wetting \cite{Balankin2000} as well as vortex motion in high-T$_{\hbox{\tiny C}}$ superconductors \cite{Blatter1994}. The theoretical frameworks \cite{Blatter1994,Chauve2000,Kolton2009} developed to model elastic interface dynamics  are therefore highly relevant for a number of real world processes which are of interest both for their fundamental properties and eventual applications.  Indeed, theoretical studies  of single interface dynamics and statics have  revealed a lot of interesting physics with predictions of universality, in-depth studies of dynamic and static {critical} exponents \cite{Huse11985,Huse21985,Kardar1985,Chauve2000,Kolton2009} and the development of now well-known interface growth equations \cite{BarabasiBook}. Magnetic systems in particular have been an ideal testing ground for these theories \cite{Lemerle1998,KrusinElbaum2001,Repain2004epl,Metaxas2007,Kim2009} since these systems can be easily probed and manipulated. 
 
A relatively recent theoretical, and more recently, experimental, playground has been developing concerning the physics of interacting interfaces in 2D systems. Theoretically, this problem has been studied via modified growth equations \cite{Barabasi1992,Majumdar2005}, Monte Carlo modeling of repulsive or non-interacting interfaces \cite{Juntunen2007,Juntunen2010} and scaling arguments \cite{Bauer2005}. Quasi-2D experimental realizations of systems containing coupled interfaces have also been conceived, ranging from interacting fluid fronts \cite{Balankin2006} to repulsive \cite{Bauer2005} and attractive \cite{Metaxas2010,SanEmeterioAlvarez2010} magnetic domain walls.

\begin{figure}
\includegraphics[width=8cm,clip=yes]{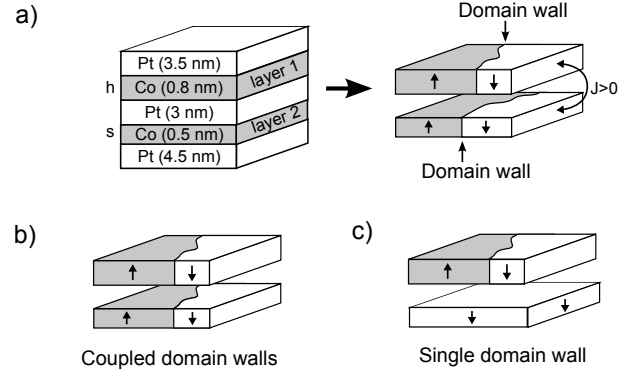}
\caption{(a) Using  magnetic layers coupled via an interlayer interaction energy, $J$, to create a model system for studying bound domain walls. Measurements of domain wall motion in this system consisted either of (b) coupled domain walls (boundaries of domains which are statically aligned in zero field) or (c) single domain walls (boundaries of domains existing in the hard layer only). 
}
\label{fig_basics1}
\end{figure}

Our work on field-driven, attractively coupled domain walls \cite{Metaxas2010} has been carried out on a system consisting of two physically separate, but magnetically coupled\cite{Moritz2004}, ultrathin ferromagnetic Co layers [Fig.~\ref{fig_basics1}(a)]. 
The ferromagnetic coupling tends to align the magnetization in the two layers. Therefore, if a domain wall is present in each Co layer [eg.~Fig.~\ref{fig_basics1}(b)], the ferromagnetic
coupling will tend to align them, acting as an attractive
interaction between the walls. This attraction not only favors a {\it static} domain wall alignment in zero-field, but can also stabilize the aligned state dynamically under an applied field\cite{Metaxas2010}. In this case, walls in the two layers are dynamically bound and move together at a common, unique velocity, despite each wall having different intrinsic velocity-field responses. These differing velocity-field responses however do mean that dynamic domain wall binding can occur only over field ranges in which wall velocities in each layer are sufficiently close, placing a limit on the fields for which bound motion can occur.  Until this work, studies of pairs of interacting interfaces in quasi-2D systems had been mostly carried out in single media. While it was already thought that domain walls in strongly coupled layers moved together \cite{Wiebel2005,Wiebel2006,Metaxas2009}, this was the first study wherein both dynamically bound domain walls and transitions between bound and unbound dynamics were directly evidenced. 

In this article, we discuss in detail a theoretical description of bound domain wall motion. The paper is outlined as follows. In Sec II we briefly give some details about the model system. 
In Sec. III we analyze how domain wall speed is affected by interlayer coupling and in
Sec. IV we study analytically the bound state regimes and discuss
the agreement between theory and experiment.
A short conclusion follows.


\section{Coupled ultrathin magnetic layers}

The experimental system shown in Fig.~\ref{fig_basics1}(a) is a magnetic multilayer consisting of two ultrathin Co layers: a magnetically hard 0.8 nm layer (layer 1) and a softer 0.5 nm layer (layer 2). The layers are ferromagnetically coupled~\cite{Moritz2004} (coupling energy $J>0$) across a 3 nm thick Pt spacer.  Seed and capping Pt layers ensure an out-of-plane magnetic anisotropy within the Co layers. Pt/Co-based films are now considered good experimental realizations of a weakly
disordered, ferromagnetic 2D Ising system, due to their anisotropy-induced out-of-plane magnetization, narrow domain walls  and intrinsic structural disorder  \cite{Lemerle1998,Metaxas2007}. This disorder has a major role in determining the velocity response $v(H)$ of a domain wall
to an external  field $H$, applied perpendicular to the film plane.

Two types of domain wall velocity measurements were carried out based upon the two domain (wall) types which could be nucleated within the multilayer. Both types of wall could be propagated under field to determine their velocity-field responses using a quasi-static magneto-optical method \cite{Metaxas2007,Metaxas2010}. (1) Coupled domain walls are the boundaries of domains existing in both layers which, in zero field, are aligned spatially with their magnetizations pointing in the same direction, as shown in Fig.~\ref{fig_basics1}(b). 
Under field, and depending on the  field amplitude, they can move together, in a dynamically bound state, or separately.  (2) Single domain walls are the boundaries of  domains existing in the hard layer only,
as illustrated in Fig.~\ref{fig_basics1}(c).  Measurements of these domain walls yield a reference velocity and a determination of the interlayer coupling.


\section{From isolated to coupled and bound domain wall dynamics}

\begin{figure}
\includegraphics[width=8cm,clip=yes]{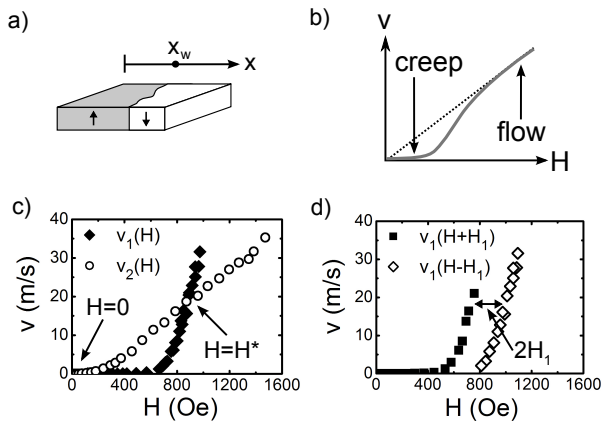}
\caption{(a) The average domain wall position is denoted $x_w$. (b) At finite temperature, walls exhibit a low field, thermally activated creep regime and a high field, dissipation-limited, linear flow regime. The two are separated by a thermally smeared depinning transition\cite{Bustingorry2008} (not labeled). (c) Experimentally obtained domain wall dynamics in layers 1 and 2 in the absence of coupling. The two curves cross at $H=0$ and $H=H^*$. (d) Domain wall dynamics in layer 1 for a coupling field, $H_1$, which reinforces the applied field $H$, ($v_1(H+H_1)$) or works against it ($v_1(H-H_1)$). }
\label{fig_basics2}
\end{figure}

Here we analyze field-velocity responses of:
i)~a single domain wall in an isolated magnetic layer,
ii)~a single domain wall in one magnetic layer coupled to a second, saturated magnetic layer, and
iii)~two coupled  domain walls, one in the hard layer and the other in the soft layer.
Domain walls will be approximated as straight lines, whose position is given by
a single number.
We begin with
a single wall  located at $x=x_w$ in an isolated ultrathin Co layer [see Fig.~\ref{fig_basics2}(a)]. The Co 
layer is  positively magnetized for $x<x_w$ and negatively magnetized for $x>x_w$.
The application of an external field $H>0$ drives the wall to the right, with the wall acquiring a
positive velocity $v(H)=dx_w/dt$. Experimental results\cite{Metaxas2007} obtained for domain wall motion in Pt/Co(0.5-0.8 nm)/Pt films
 show that  $v(H)$ is characterized by two distinct regimes at room temperature  (creep and flow) which were theoretically predicted\cite{Blatter1994,Chauve2000} and are sketched in the schematic of Fig.~\ref{fig_basics2}(b). Domain walls exhibit flow motion at high fields for which $v\propto H$.
However, below a layer-dependent critical depinning field $\Hdep$ 
(generally on the order of a few hundred Oersted\cite{Metaxas2007}), disorder-induced pinning effects become significant and the walls exhibit thermally activated creep\cite{Lemerle1998}. Within this latter regime, $v(H)$ has the form
\be
v(H) = v_0 \exp\left[ -{U_C\over k_B T}\left({\Hdep\over H}\right)^{1/4}\right] , \label{creep_equation}
\ee
where the exponential factor $U_C/k_B T$ is the ratio between the typical pinning energy and the thermal
 energy.
The exponent $1/4$
is a universal exponent, characteristic of the dynamics of  a one dimensional interface
in a 2D weakly disordered medium.

Films with different thicknesses have different microscopic parameters and
disorder strengths. As a result, they have different $v(H)$ characteristics\cite{Metaxas2007}, as attested
by the experimental velocity-field curves for domain walls in the two layers in the absence of coupling [Fig.~\ref{fig_basics2}(c)]. 
However, pairs of such curves often intersect
at two points: $H=0$ and $H=H^*$ ($H^*\simeq 860$~Oe for our system). The first crossing point is 
universal, because the velocity $v(f)$ of any isolated interface in
response to a generalized force $f$ (here, $f=H$) is always expected to vanish
for vanishing $f$.
The second crossing point is less trivial and arises because domain walls in thicker Co layers generally have a lower creep velocity but a higher flow velocity than walls in thinner layers.

In the remainder of this article, we shall use $v_1(H)$ to refer to the domain wall velocity in the hard layer and $v_2(H)$ to that
in the soft layer. 
If the two films are not coupled, it is clear  that walls will propagate independently, with
$v_2>v_1$ for $H<H^*$ and $v_1>v_2$ for $H>H^*$ [Fig.~\ref{fig_basics2}(c)]. The question we are now going to consider is the following: 
What is the effect of interlayer coupling on domain wall velocities $v_{1,2}(H)$ and domain wall binding phenomena?

Before considering 
coupling between domain walls, let us consider the simpler case of a single domain wall 
in layer $i=1,2$, interacting with a uniformly magnetized layer $k=2,1$ (see, for example, Fig.~\ref{fig_basics1}(c)). The interlayer coupling
$J$ induces an effective coupling field $h_i$, given by~\cite{Grolier1993-2} 
\be
h_i=m_k\frac{J}{\Ms^i t_i}\equiv m_k H_i ,
\label{eq_coupling_energy}
\ee
where $\Ms$ is the saturation magnetization, $t$ is the layer thickness, and $m=\pm 1$ is the magnetization
orientation. $h_i$ adds to the external field $H$ and also drives the domain wall\cite{Fukumoto2005,Metaxas2008}, in turn allowing for a simple experimental determination of $H_i$. To determine $H_1$, domain wall velocities in the hard layer were measured while keeping the soft layer magnetically saturated. Through control of $m_2$ and/or $H$, it was possible to determine wall velocities with $h_1$ either opposing or reinforcing the applied field. We denote these data sets $v_1(H-H_1)$ and $v_1(H+H_1)$ respectively. Plotted in Fig.~\ref{fig_basics2}(d), the two data sets are separated by $2 H_1$, allowing a determination of $H_1=120$ Oe and $v_1(H)$ [ie.~no coupling, see Fig.~\ref{fig_basics2}(c)].

The corresponding coupling field and isolated wall dynamics for layer 2 were determined in a different manner. $H_2=220$ Oe could be easily found using Eq.~(\ref{eq_coupling_energy}), which
gives $H_1 \Ms^1 t_1 = H_2 \Ms^2 t_2$ ($\Ms^{1,2}$ are known \cite{Metaxas2010}). 
Unfortunately, we were not able to nucleate a domain in the soft layer while keeping the
hard layer in a single domain state and so $v_2(H)$ had to be
measured using a Co(0.5 nm) layer in a less strongly coupled Pt/Co(0.5 nm)/Pt(4 nm)/Co(0.8 nm)/Pt film \cite{Metaxas2008}.

Now, let's turn to dynamics of coupled walls [Fig.~\ref{fig_basics1}(b)].
The experimental determination of the coupled walls is as follows.
(i)~Two aligned domain walls, at a common position
$x_1(0)=x_2(0)$, are nucleated. (ii)~A magnetic field pulse, $H$, is applied for a time $T$, under which
walls move to positions $x_1(T),x_2(T)$. (iii)~The new wall positions are quasi-statically determined\cite{Metaxas2007,Metaxas2010} from Kerr microscopy images. 

While $x_1(T)=x_2(T)$ for dynamically bound walls, $x_1(T)\ne x_2(T)$ for unbound walls since the walls separate during their motion. However, the time interval between steps (ii) and (iii) is large enough to allow the separated walls to relax back to an aligned state under the action
of effective coupling fields ($H=0$ for $t>T$). Since $v_2(H_2)/v_1(H_1)\approx 10^{10}$, if $x_1(T)\ne x_2(T)$, pre-imaging relaxation of the soft layer wall gives: 
$x_2^{\hbox{\tiny imaged}}
=x_1(T)$. Therefore, the experimental technique yields either the true bound wall displacement (and subsequently the bound velocity) or the  hard wall displacement (and therefore the hard layer wall velocity) when the walls are unbound. 

\begin{figure}
\includegraphics[width=7cm,clip=on]{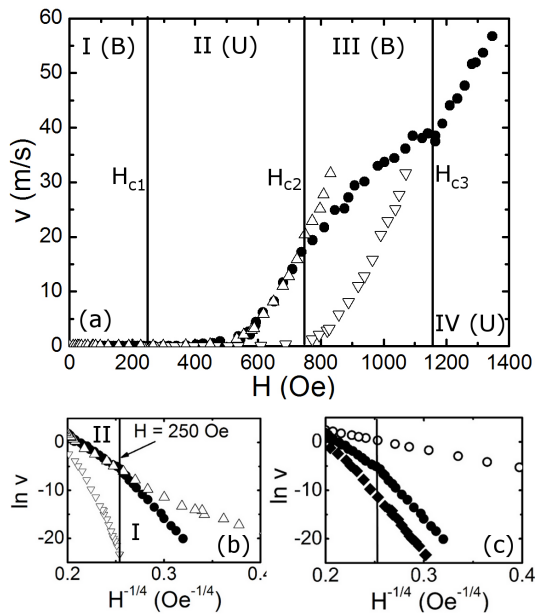}
\caption{ (a and b) Experimentally obtained coupled wall velocity, $\vc(H)$ ($\bullet$) plotted with  hard domain velocities in the presence of a positively saturated soft layer, $v_1(H+H_1)$ ($\bigtriangleup$) and a negatively saturated soft layer, $v_1(H-H_1)$ ($\bigtriangledown$). Field regions in which bound (B) and unbound (U) coupled wall dynamics are labeled with roman numerals. Vertical solid lines represent the region limits ($H_{c(1,2,3)}$). (c) $\vc(H)$ ($\bullet$) compared to hard and soft layer domain wall creep velocities in the absence of coupling, $v_1(H)$ ($\blacklozenge$) and $v_2(H)$ ($\circ$) respectively.}
\label{fig_exp}
\end{figure}

In the unbound state, the hard layer velocity (and therefore the experimentally determined velocity of the coupled walls), will be that observed for hard layer walls under a field $H\pm H_1$ [Fig.~\ref{fig_basics2}(d)] since the walls in the two layers are not aligned: $+H_1$ if the hard layer wall trails the soft layer wall and $-H_1$ if the hard layer wall leads the soft layer wall [Eq.~(\ref{eq_coupling_energy})]. This is an important point, as it allows us to  identify the field ranges over which $\vc(H)$ (the experimentally obtained coupled wall velocity) corresponds to unbound motion. The unbound (U) and bound (B) regions are labeled in Fig.~\ref{fig_exp}(a) in which $\vc(H)$ is plotted together with $v_1(H\pm H_1)$ to allow a direct comparison. This allows us to easily locate the three critical fields, $H_{c1,2,3}$, which separate bound and unbound states [see vertical lines
in Figs.~\ref{fig_exp}(a,b)]:
$H_{c1}\approx 250$Oe, $H_{c2}\approx 750$Oe, and $H_{c3}\approx 1150$Oe.

\begin{figure*}
\includegraphics[width=15cm,clip=yes]{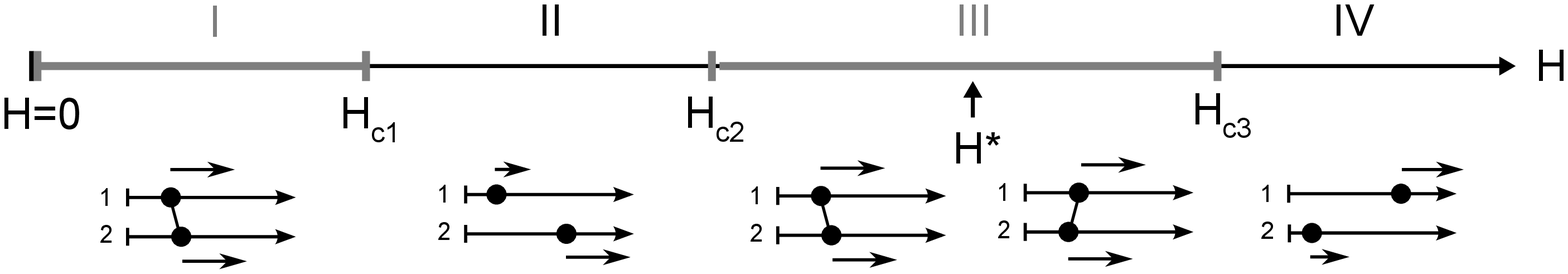}
\caption{This figure shows the different field regimes.
In the unbound states (thin, black lines), walls in layer 1 and 2 move at different speeds, the leading wall
at a higher velocity.
In the bound states (thick, grey lines), walls move at the same speed. The distance between
walls is therefore constant. The leading wall is the wall with the higher velocity, 
in the absence of coupling.
}
\label{fig_regimes}
\end{figure*}

We first consider the unbound field ranges. In region II, $H_{c1}<H<H_{c2}$ of Fig.~\ref{fig_exp}(a), 
$v_2(H)\gg v_1(H)$ [see Fig.~\ref{fig_basics2}(c)], so that the soft domain wall leads
and the distance $(x_2 - x_1)$ between walls is positive and large. The soft wall
is so far ahead of the hard wall that the latter moves under the action of a
positively saturated soft layer. When $H> H_{c3}$, the situation is reversed: $v_1(H)\gg v_2(H)$ [see Fig.~\ref{fig_basics2}(c) again]. The hard domain 
wall leads and the soft wall is so far behind it that the hard wall moves under
the action of a negatively saturated soft layer. A schematic of these regimes is shown in Fig.~\ref{fig_regimes}.

While we can compare $\vc(H)$ to $v_1(H\pm H_1)$ to obtain values for the region limits $H_{cj}$ ($j=1,2,3$), these values can also be evaluated from the experimentally obtained velocity data  in Fig.~\ref{fig_basics2}(c) and the $H_{1,2}$ values.
Before moving on to analytical and numerical modeling results, we explain how this is done using a simple graphical method.

In regime II, the walls in each layer move separately with the soft wall leading. This can be sustained
only if 
\be
v_2(H-H_2)>v_1(H+H_1).
\label{eq_regimeI}
\ee
Therefore, it is straightforward to define
the critical fields $H_{c1}$ and $H_{c2}$ through the equation
\be
v_2(H-H_2) = v_1(H+H_1).
\label{eq_Hc12}
\ee
This equation can be solved graphically using the data in Fig.~\ref{fig_basics2}(c) 
to give $H_{c1}\sf\approx 260$Oe and
$H_{c2}\sf\approx 600$Oe,
where the superscript means that critical field values have been determined by the
single, isolated, film velocities $v_{1,2}(H)$.
Similarly, in regime IV, the walls are unbound again but with the hard wall leading. This can be sustained only if 
\be
v_1(H-H_1)>v_2(H+H_2).
\label{eq_regimeIII}
\ee
The equation
\be
v_2(H+H_2) = v_3(H-H_1)
\label{eq_Hc3}
\ee
now has only one solution, which gives the lower limit of regime IV: 
$H_{c3}\sf\approx 1050$Oe.
The value $H_{c1}\sf$ compares quite well with $H_{c1}$, obtained from a visual inspection
of the $\vc(H)$ and $v_1(H\pm H_1)$ data above.
The bounds $H_{c2}\sf,H_{c3}\sf$ of region III do not compare so well with
$H_{c2},H_{c3}$. We will comment on that in Sec.~\ref{sec_high_field}.

Having considered regions II and IV, we can now turn to the remaining regions, regions I and III, which are located around the crossing fields, $H=0$ (region I) and
$H=H^*$ (region III). In these field regions, the two walls cannot move separately
at different speeds, because neither Eq.~(\ref{eq_regimeI}) nor Eq.~(\ref{eq_regimeIII})
is satisfied. In the following Section we argue that in this case a bound state
arises, for which the common domain wall speed depends on $v_i(H)$ in a non-trivial way.


\section{Numerical and analytical results for bound states}

\subsection{One-dimensional model for wall dynamics and numerical results}

In the following we want to introduce a minimal, one-dimensional model, which can explain
the rising of dynamically bound states and gives quantitative expressions for the common 
speed of two coupled walls. Each domain wall is approximated
by its average position $x_i(t)$, $i=1,2$ [Fig.~\ref{fig_basics2}(a)]. A total field $(H+\bar H_i(x))$ acts
on the $i-$th wall. It is the sum of the external field $H$ and the coupling field $\bar H_i(x)$,
which depends on the distance $x=x_2-x_1$ between walls.
We expect that the coupling field $\bar H_i$ is equal to $\pm H_i$, if the two walls are well separated,
with the plus (minus) sign applying for the trailing (leading) wall.
It is useful to make the following general assumption for the coupling fields:
\be
\bar H_1(x) = H_1 f(x) ~~~~ 
\bar H_2(x) = -H_2 f(x) ,
\label{coupling_fields}
\ee
where $f(x)$ is an unspecified odd function,
interpolating between $-1$ and $+1$, as $x$ varies from negative to positive values.
Each wall moves with the velocity $v_i(H+\bar H_i(x))$. A bound state corresponds to motion with
\be
v_1 (H - H_1 f(x)) = v_2 (H + H_2 f(x))
\label{eq_bound_motion}
\ee
for some value $x$, corresponding to the constant distance between walls. 
If Eq.~(\ref{eq_bound_motion}) has no solution, it means that the
walls are unbounded (and therefore separated) either with the wall in the hard layer leading ($v_1(H-H_1) > v_2(H+H_2)$) or with the wall in the soft layer leading ($v_2(H-H_2) > v_1(H+H_1)$).

If we define the ratio $\alpha=H_2/H_1$ between coupling fields, we easily find that the solution 
$x=x_0$ of Eq.~(\ref{eq_bound_motion}),
\be
v_1 (H - H_1 f(x_0)) = v_2 (H + \alpha H_1 f(x_0))
\ee
has the form 
\be
H_1 f(x_0) = G(H,\alpha)
\label{eq_G}
\ee
and the common speed $v_b(H)$ of bound motion is
\be
v_b(H) = v_1 (H - G(H,\alpha)) = v_2 (H+\alpha G(H,\alpha)) .
\label{eq_vbound}
\ee

Therefore, the specific form of the function $f(x)$ is irrelevant to determine the velocity
of bound motion: the speed depends only on the external field $H$ and the ratio $\alpha$ between
coupling fields. Different forms of $f(x)$ give different equilibrium distances $x_0$, but the
same common velocity~\cite{Note1}.

\begin{figure}[tbp]
\includegraphics[width=8cm,clip=yes]{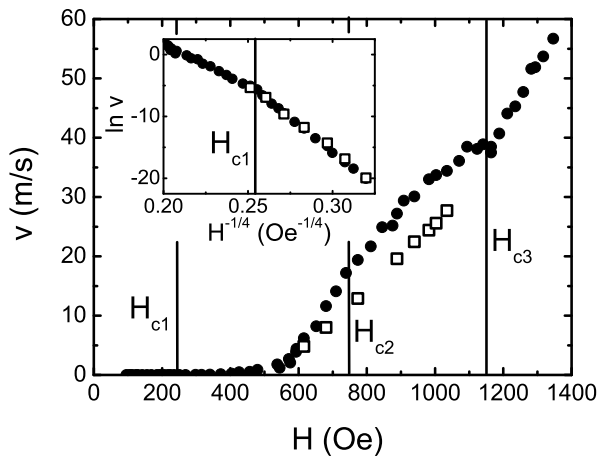}
\caption{Comparison between experimentally obtained coupled wall dynamics, $\vc(H)$ ($\bullet$), 
and the theoretically calculated bound state velocity, $v_b(H)$ ($\square$). }
\label{fig_exp-theory}
\end{figure}

We can now solve Eq.~(\ref{eq_bound_motion}) using experimental data for single wall motion ($v_1(H)$ and $v_2(H)$, Fig.~\ref{fig_basics2}(c)). 
This way, our theory provides the velocity of bound states without free parameters.
Results are shown in Fig.~\ref{fig_exp-theory}. Comparison with experimental data is very
satisfying for the low field bound state regime, with modest quantitative agreement in the
high field bound state regime. In the next Sections we are going to discuss
both regimes in more detail and derive analytical expressions describing the bound dynamics. 

\subsection{The low field bound state regime}

In the creep regime, analytical expressions are available for the wall velocities in the uncoupled
case [see~Eq.~(\ref{creep_equation})],
\bse
\label{eq_creep}
\be
v_1(H) = v_1^0 \exp\left[-\left({a_1\over H}\right)^{1\over 4}\right] 
\ee
\be
v_2(H) = v_2^0 \exp\left[-\left({a_2\over H}\right)^{1\over 4}\right] , 
\ee
\ese
where experimental values for $v_i^0$ and $a_i$ are given in Table \ref{tab_exp}.

\begin{table}[thbp]
\begin{tabular}{|c|c|c|c|}
\hline
               & Coupling field & exponent & prefactor \\
               & (Oe) & (Oe$^{1/4}$) & $\ln$(m/s) \\
\hline
Hard (1) layer & $H_1=120$    & $a_1^{1\over 4} = 225.2$ & $\ln v_1^0 = 45.3$ \\ 
\hline
Soft (2) layer & $H_2=220$    & $a_2^{1\over 4} = 40.1$ & $\ln v_2^0 = 10.8$ \\ 
\hline
Bound creep & $-$    & $a_b^{1\over 4} \approx 202$ & $\ln v_b^0 \approx \ln v_1^0$ \\ 
\hline
\end{tabular}
\caption{Experimental values for the coupling field magnitudes and  parameters 
for uncoupled domain wall creep dynamics [Eqs.~(\ref{eq_creep})] as well as bound creep [Eq.~(\ref{eq_vb})].} 
\label{tab_exp}
\end{table}

We are now going to prove that, in the limit $H\to 0$, Eq.~(\ref{eq_bound_motion}) has
a solution which describes a \textit{bound creep} motion such that the common domain wall velocity is given by
\be
v_b(H) = v_b^0 \exp\left[-\left({a_b\over H}\right)^{1\over 4}\right].
\label{eq_vb}
\ee

In order to reduce the notation, let us introduce the quantity $c=f(x)$, which varies in 
the interval $(-1,+1)$.  We have to solve Eq.~(\ref{eq_bound_motion}), which
using Eqs.~(\ref{eq_creep}), can be written as
\be
v_1^0 \exp\left[-\left({a_1\over H-cH_1}\right)^{1\over 4}\right] =
v_2^0 \exp\left[-\left({a_2\over H+cH_2}\right)^{1\over 4}\right] .
\label{eq_1dmodel}
\ee

It is clear that walls must move with a positive velocity, if the external field $H$ is 
positive.
This requires the sign of the total driving fields
$H\pm c H_i$ to be the same as the sign of $H$, which demands that
$c$ vanishes in the limit $H\to 0$.
Therefore, we use a small $H$ expansion
\be
c= c_0 H + c_1 H^{1+\gamma} ,
\label{c_expansion}
\ee
where the value of $\gamma$ will be found  below, while
it is straightforward that the leading term is linear. In fact, 
if $c$ vanishes faster than linearly, 
the coupling would not have effect in the limit $H\to 0$ and a bound state would be impossible
for small $H$. On the other hand,
if $c$ vanishes slower than linearly, $H\pm c H_i$ cannot both have the same sign as $H$.
In conclusion, using (\ref{c_expansion}) we can rewrite Eq.~(\ref{eq_1dmodel}) as
\begin{widetext}
\be
v_1^0 \exp\left\{-\left({a_1\over H[1-H_1(c_0+c_1 H^\gamma)]}\right)^{1\over 4}\right\} 
=v_2^0 \exp\left\{-\left({a_2\over H[1+H_2(c_0+c_1 H^\gamma)]}\right)^{1\over 4}\right\} 
\equiv  v_b^0 \exp\left[-\left({a_b\over H}\right)^{1\over 4}\right],
\label{eq_extended}
\ee
where we have used the fact that the common speed must have the form (\ref{eq_vb}).
Equation (\ref{eq_extended}) can be rewritten as
\be
v_1^0 \exp\left[-\left({a_1\over H(1-c_0 H_1)}\right)^{1\over 4}
\left( 1- {c_1H_1\over 1-c_0H_1}H^\gamma\right)^{-1/4} \right]
=v_2^0 \exp\left[-\left({a_2\over H(1+c_0 H_2)}\right)^{1\over 4}
\left( 1+ {c_1H_2\over 1+c_0H_2}H^\gamma\right)^{-1/4} \right] ,
\ee
which can be approximated, in the limit of vanishing $H$, as
\be
v_1^0 \exp\left[-\left({a_1\over H(1-c_0 H_1)}\right)^{1\over 4}
\left( 1+ {{1\over 4}c_1H_1\over 1-c_0H_1}H^\gamma\right) \right]
=v_2^0 \exp\left[-\left({a_2\over H(1+c_0 H_2)}\right)^{1\over 4}
\left( 1- {{1\over 4}c_1H_2\over 1+c_0H_2}H^\gamma\right) \right] . 
\label{eq_expansion}
\ee

If we take the logarithm of both sides, we get
\be
\ln v_1^0 -\left({a_1\over H(1-c_0 H_1)}\right)^{1\over 4}
\left( 1+ {{1\over 4}c_1H_1\over 1-c_0H_1}H^\gamma\right)
=\ln v_2^0 -\left({a_2\over H(1+c_0 H_2)}\right)^{1\over 4}
\left( 1- {{1\over 4}c_1H_2\over 1+c_0H_2}H^\gamma\right)
\equiv \ln v_b^0 -\left( {a_b\over H} \right)^{1/4} ,
\label{eq_Hexp}
\ee
\end{widetext}
which is put in the form of Eq.~(\ref{eq_vb}).

The equality in Eq.~(\ref{eq_Hexp}) requires, to leading order in $H$, that
\be
\left({a_1\over H(1-c_0 H_1)}\right)^{1\over 4}
=\left({a_2\over H(1+c_0 H_2)}\right)^{1\over 4} 
\equiv \left( {a_b\over H} \right)^{1/4} ,
\label{eq_leading}
\ee
that is to say
\be
{a_1\over 1-c_0 H_1} = {a_2\over 1+c_0 H_2} \equiv a_b ,
\label{eq_a_b}
\ee
which gives
\be
c_0 = {a_2 - a_1\over a_1 H_2 + a_2 H_1}.
\label{eq_c_0}
\ee

If we replace Eq.~(\ref{eq_a_b}) in Eq.~(\ref{eq_Hexp}), we get
\be
\ln v_1^0 - {{1\over 4}a_b^{1/4}c_1H_1\over 1-c_0H_1}H^{\gamma-{1\over 4}}
=\ln v_2^0 + {{1\over 4}a_b^{1/4}c_1H_2\over 1+c_0H_2}H^{\gamma-{1\over 4}}
\equiv \ln v_b^0
\label{eq_2ndorder}
\ee
which has a solution for $c_1$ only if $\gamma={1\over 4}$:
\be
c_1 = \frac{
\ln \left( {v_1^0\over v_2^0} \right)
}{
{1\over 4} a_b^{5/4} \left( {H_1\over a_1} + {H_2\over a_2} \right)
} .
\ee
Replacing $c_1$ in the left or middle expression of (\ref{eq_2ndorder}), we get
\be
\ln v_b^0 = \ln v_1^0 - {a_2 c_0 H_1\over a_2 -a_1} \ln\left({v_1^0\over v_2^0}\right) .
\ee

\begin{figure}[tbp]
\includegraphics[width=7cm,clip=on]{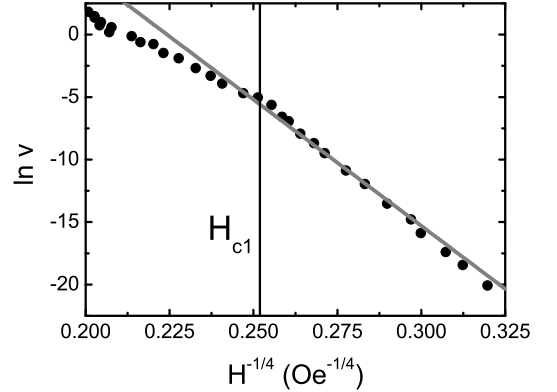}
\caption{Comparison between experimentally obtained $\vc(H)$ ($\bullet$) 
and analytical bound state velocity (full line) [Eq.~(\ref{eq_vb})  using creep parameters given in Eq.~(\ref{res_1dmodel})]. 
}
\label{fig_compare_analytics}
\end{figure}

Using experimental values for separated domain wall velocities, see Tab.~\ref{tab_exp},
we find that $a_2/a_1 \simeq 10^{-3}$ and $H_1 < H_2$, so that 
(see Eq.~(\ref{eq_c_0})), $c_0\approx -1/H_2$.
A negative $c_0$ means that walls move at the same speed as a bound state,
with the soft wall leading (see Fig.~\ref{fig_regimes}). This is expected,
because in the uncoupled case, $v_2(H)>v_1(H)$ for small $H$.
Finally, we get 
\be
a_b^{1/4} \simeq 202\hbox{Oe$^{1/4}$~~~~~~~}\ln v_b^0 \simeq \ln v_1^0 .
\label{res_1dmodel}
\ee
$a_b^{1/4}$ is closer to $a_1^{1/4}$ than $a_2^{1/4}$ as previously noted\cite{Metaxas2010} and seen in Fig.~\ref{fig_exp}(c).
Notably, we can substitute the above creep parameters into Eq.~(\ref{eq_vb}) to have a complete analytic expression for the bound state velocity which compares well to the low field $\vc(H)$ data below $H_{c2}$ [see Fig.~\ref{fig_compare_analytics}(a)]. 

\subsection{High field bound state regime}\label{sec_high_field}

Let us now consider the high field bound state around $H^*$.
In this regime,
comparison between the one-dimensional model and experimental results show only modest agreement. 
Even if our theory correctly anticipates the
existence of a bound state regime around $H=H^*$, the agreement between observed
($H_{c2,3}$) and predicted ($H_{c2,3}\sf$) limit field values is not perfect. Furthermore, 
the theory ($v_b(H)$)
underestimates the experimental ($\vc(H)$) 
bound state velocity, $v_b(H) < \vc(H)$, for $H_{c2}<H<H_{c3}$. Below, we discuss these details and, 
in particular, why the experimental  bound state velocity at $H^*$, $\vc(H^*)\simeq 24.5$m/s, is significantly larger than 
$v_1(H^*)=v_2(H^*)=v^*\simeq 18$m/s.
In Appendix \ref{app_high_field} we also give an analytical approximation for the bound state
velocity in the high field regime.
Finally, it is worth mentioning that the constant distance $x_0=x_2-x_1$ between
the soft and the hard walls in the bound regime, is positive for $H<H^*$ and negative
for $H>H^*$ (see Fig.~\ref{fig_regimes}), because the {\it leading} wall in the bound regime
is the wall with the highest speed in the absence of coupling.

Now, let us discuss the disagreement between our theory and experimental results
in the high field bound regime.
There are three main possibilities to explain this: (1) our coupling model is inadequate, (2) the data used for $v_2(H)$ is not representative of the true $v_2(H)$ in this system or (3) the use of the experimental $v_{1,2}(H)$ data is not valid for the high field limit.

(1) In Appendix \ref{appendix_coupling}, we discuss two modifications to the coupling: a dipolar coupling (additional $g(x)$ term in Eqs.~(\ref{coupling_fields}): see Eqs.~(\ref{eq_app1})) due to strong stray fields at the domain edges\cite{Baruth2006} and the use of differing $f_1(x)$ and $f_2(x)$ functions in Eqs.~(\ref{coupling_fields}) (see Eqs.~(\ref{eq_app2})). However, both modifications still lead to $v_b(H^*)=v^*$. Furthermore, since the low field bound regime is well reproduced
using only the exchange field, it is questionable to make Eqs.~(\ref{coupling_fields}) 
more complicated. One might also consider the case in which $f(x)$ is not continuous. For example,
we might have a step function, $f(x)=-1$ for $x<0$ and $f(x)=+1$ for $x>0$. This implies
that the bound state is not characterized by a constant distance between walls, but by a
continuous interchange between the walls. However, this neither solves the issue surrounding $v_{b}(H^*)\ne v^*$, nor the discrepancy between $H_{c2,3}$ and $H_{c2,3}\sf$.

(2) As explained earlier, $v_2(H)$ was not measured in this multilayer but rather in a similar one with an equivalent Co(0.5 nm) layer. Using this data, we see that in the vicinity of $H^*$, walls in layer 2 exhibit flow motion wherein $v_2=mH$ with $m\approx 0.022$ ms$^{-1}$Oe$^{-1}$. There can be some sample to sample variability however and previous measurements on a single layer Pt/Co(0.5 nm)/Pt film\cite{Metaxas2007} yielded $m\approx 0.027$ ms$^{-1}$Oe$^{-1}$. Using this $m$ value to model dynamics in layer 2 for the purpose of determining $v_b(H)$ at high field changes $H^*$, which is now equal
to $H^*=910$Oe. The new $m$ value
 improves the consistency between our calculated $H_{c2}\sf$ and $H_{c3}\sf$ values (700 Oe and 1070 Oe, respectively) as compared to the experimental values, $H_{c2}$ and $H_{c3}$ (750 Oe and 1150 Oe, respectively). However, the newly calculated value of $v_b^*$ (24.5 m/s) remains too low with respect
to the experimental value $\vc(H^*=910\;\hbox{Oe})\simeq 29$m/s~\cite{Note2}.
Note that the film measured in Ref.~\onlinecite{Metaxas2007} also had a slightly lower $a^{1/4}$ value (35.1) as compared to $a_2^{1/4}$ (40.1, see Table 1), however this has little effect on the predicted bound dynamics since they are dominated by the larger $a_1^{1/4}$ value. 

(3) Finally, our approach, which works well at low field, may not actually be appropriate at high field where wall dynamics are intrinsically different. At low field, wall motion is thermally activated over field-dependent energy barriers. In contrast, at high field, wall motion is, to a large extent, determined by the internal structure of the wall (and associated internal dynamics) \cite{Schryer1974,Mougin2007} which can actually be  modified by interlayer coupling \cite{Bellec2010}. As such, experimentally obtained, isolated single wall velocities $v_1(H)$ and $v_2(H)$ may not be the
appropriate building blocks to be combined to calculate $v_b(H)$, as we did in Eq.~(\ref{eq_vbound}).

\section{Conclusion}

Exchange coupled Pt/Co layers represent an ideal model experimental system in which to study the interesting problem of coupled interfaces moving through physically separate, but coupled, media. Here we have detailed the principles behind this system and presented both numerical and analytical models of bound domain wall motion which compare well with experiment\cite{Metaxas2010}. Most notably, we derive an analytical model with no free parameters which describes bound creep.  While we have concentrated on a one dimensional model we hope out results will inspire others to apply micromagnetic\cite{Martinez20072,Bellec2010} or interface models\cite{Kolton2005,Kolton2009} to this problem.


\appendix

\begin{acknowledgments}
The authors wish to thank B.~Rodmacq and V.~Baltz for useful discussions and for providing samples.
 P.P., P.J.M. and R.L.S.  acknowledge support from the Australian Research Council and the Italian Ministry of Research (PRIN 2007JHLPEZ). P.J.M. acknowledges support from an Australian Postgraduate Award and a Marie Curie Action (MEST-CT-2004-514307). P.J.M., R.L.S. and J.F. also received support from the French-Australian Science and Technology (FAST) Program.
\end{acknowledgments}

\section{Additional and modified coupling}\label{appendix_coupling}

A more general expression of
Eqs.~(\ref{coupling_fields}) which includes dipolar interactions might be
\be
\bar H_1 = H_1 f(x) + D_1 g(x) ~~~~
\bar H_2 = -H_2 f(x) - D_2 g(x)
\label{eq_app1}
\ee
where 
$g(x)=\ln [(d^2 + x^2)/d^2]$ accounts for dipolar coupling, $d=3$ nm being the separation between
the hard and the soft layer. A simple calculation\cite{Baruth2006} can show that in the vicinity of the domain walls, dipolar fields can potentially be larger than $H_{1,2}$. However, as discussed in Sec.~\ref{sec_high_field}, the good agreement between theoretical and experimental results at low field suggests that it is $H_{1,2}$ which determine the  bound state's stability.

An alternative generalization of Eqs.~(\ref{coupling_fields}) is to make $f(x)$ different for
the two films,
\be
\bar H_1 = H_1 f_1(x)  ~~~~
\bar H_2 = -H_2 f_2(x) .
\label{eq_app2}
\ee
This might mean, e.g., writing 
\be
f_i(x)=\tanh \left({x\over \Delta_i}\right),
\ee
with $\Delta_1\ne\Delta_2$, as is expected for layers with differing thicknesses\cite{Metaxas2007}.

However, both of these approaches yield $v_b(H^*)=v^*$ since $g(0)=0$ and $f_1(0)=f_2(0)=0$. 

\section{Analytical approximation for the high field bound state}
\label{app_high_field}

In the high field regime, the walls are no longer in the creep regime and Eqs.~(\ref{eq_creep})
cannot be used. Instead, we can assume a simple linear approximation\cite{Lemerle1998,Metaxas2007} in the proximity of $H=H^*$,
\be
v_i(H) = v^* + \bai (H-H^*) ,
\ee
where $v^*=v_i(H^*)$ and $\bai\equiv d v_i/dH|_{H=H^*}$.
It can be easily shown that the solution $x=x_0$ of Eq.~(\ref{eq_bound_motion}) satisfies the relation
\be
f(x_0) = -{(\bau-\bad)(H-H^*)\over \bau H_1 + \bad H_2},
\ee
so that 
\be
v_b(H) = v^* + \bab (H-H^*)
\ee
with
\be
\bab = {\bau\bad (H_1 + H_2)\over \bau H_1 + \bad H_2}.
\ee

Therefore, in the proximity of $H=H^*$, the common speed in the high field bound regime is linear, 
with a slope $\bab$ which is in between $\bau$ and $\bad$:
\be
\bad < \bab < \bau .
\ee

Using the fitting values $\bau\simeq 0.025$ and $\bad\simeq 0.12$, we find $\bab\simeq 0.035$,
so that
\be
v_b(H) \simeq 18 + 0.035 (H-850) ,
\ee
with $H$ expressed in Oersted and the speed in meters per second.

\end{document}